\documentstyle[12pt,aasms4]{article}

\begin{document}

\title{Quasi-Local Evolution of the Cosmic Gravitational Clustering in
        Halo Model}

\author{Long-Long Feng\altaffilmark{1,2} and
                      Li-Zhi Fang\altaffilmark{3}  }

\altaffiltext{1}{Center for Astrophysics, University of Science
and Technology of China, Hefei, Anhui 230026,P.R.China}
\altaffiltext{2}{National Astronomical Observatories, Chinese
Academy of Science, Chao-Yang District, Beijing, 100012, P.R.
China}
\altaffiltext{3}{Department of Physics, University of Arizona,
Tucson, AZ 85721}

\begin{abstract}

We show that the nonlinear evolution of the cosmic gravitational clustering
is approximately spatial local in the $x$-$k$ (position-scale) phase space
if the initial perturbations are Gaussian. That is, if viewing the mass
field with modes in the phase space, the nonlinear evolution will cause
strong coupling among modes with different scale $k$, but at the same
spatial area $x$, while the modes at different area $x$ remain
uncorrelated, or very weakly correlated. We first study the quasi-local
clustering behavior with the halo model, and demonstrate that the
quasi-local evolution in the phase space is essentially due to the
self-similar and hierarchical features of the cosmic gravitational clustering.
The scaling of mass density profile of halos insures that the coupling
between $(x-k)$ modes at different physical positions is substantially
suppressed. Using high resolution N-body simulation samples in the LCDM
model, we justify the quasi-locality with the correlation function
between the DWT (discrete wavelet transform) variables of the cosmic
mass field. Although the mass field underwent a highly non-linear
evolution, and the DWT variables display significantly non-Gaussian
features, there are almost no correlations among the DWT variables
at different spatial positions. Possible applications of the
quasi-locality have been discussed.

\end{abstract}


\section{Introduction}

The large scale structure of the universe was arisen from initial
fluctuations through the nonlinear evolution of gravitational
instability. Gravitational interaction is of long range, and therefore,
the evolution of cosmic clustering is not localized in physical space. The
typical processes of cosmic clustering, such as collapsing and falling
into potential wells, the Fourier mode-mode coupling and the merging
of pre-virialized dark halos, are generally {\it non-local}. These processes
lead to a correlation between the density perturbations at different
positions,
even if the perturbations at that positions initially are statistically
uncorrelated. For instance, in the Zel'dovich approximation (Zel'dovich 1970),
the density field $\rho({\bf x}, t)$ at (Eulerian) comoving position
${\bf x}$ and time $t$ is determined by the initial perturbation at
(Lagrangian) comoving position, ${\bf q}$, plus a displacement ${\bf S}$:
\begin{equation}
{\bf x}({\bf q}, t)= {\bf q} + {\bf S}({\bf q}, t).
\end{equation}
The displacement ${\bf S}({\bf q}, t)$ represents the effect of density
perturbations on the trajectories of self-gravitating particles. The
intersection
of particle trajectories leads to a correlation between mass
fields at different spatial positions. Thus, the gravitational clustering
is non-local even in weakly non-linear regime.

On the other hand, spatial locality has been employed in the Gaussianization
technique for recovery of the primordial power spectrum
(Narayanan \& Weinberg 1998).
Underlying this algorithm is to assume that the relation between the
evolved mass field and
the initial density distribution is local, i.e. the high(low) initial density
pixels will be mapped into high(low) density pixels of the evolved field
(Narayanan \& Weinberg 1998). Obviously, the localized mapping is difficult
in reconciling the initially Gaussian field with the coherent non-linear
structures, such as halos with scaling behavior. It has been argued that
the locality assumption may be a poor approximation to the
actual dynamics because of the non-locality of gravitational evolution
(Monaco \& Efstathiou 2000). Nevertheless, the localized mapping is found
to work well for reconstructing the initial mass field and power spectrum from
transmitted flux of the Ly$\alpha$ absorption in QSO spectra (Croft et al.
1998.) These results, joint with the data of WMAP, have been used to
determine the cosmological parameters (Spergel et al 2003). However, the
dynamical origin of the locality assumption remains a problem. It is still
unclear under which condition the localized mapping is a good approximation.

This problem has been studied in weakly non-linear regime
under the Zel'dovich approximation. The result
showed that the cosmic gravitational clustering evolution is spatially
quasi-localized in phase ($x$-$k$) space made by the DWT decomposition
(Pando, Feng \& Fang 2001). In this approach, each perturbation mode
corresponds to a cell in the $x$-$k$ space, ($x$ to $x+\Delta x$,
$k$ to $k + \Delta k$) with $\Delta x\Delta k =2\pi$, and the density
perturbation of the mode $(x,k)$ is $\delta(k, x)$. They demonstrated
that, in the
Zel'dovich approximation, if the initial density perturbations in each
cells are statistically uncorrelated, i.e.
$\langle \delta_0(k_1,x_1)\delta_0(k_2,x_2)\rangle \propto
\delta^K_{k_1,k_2}\delta^K_{x_1.x_2}$, where $\delta^K$ denotes for
Kronecker delta
function, the evolved $\delta(k_1,x_1)$ and $\delta(k_2,x_2)$ will keep
approximately spatially uncorrelated always,
$\langle \delta(k_1,x_1)\delta(k_2,x_2)\rangle\propto \delta^K_{x_1.x_2}$,
which is just what we call spatial quasi-locality in the dynamics of cosmic
clustering.
The spatial quasi-locality implies a significant local mode-mode coupling
(different scales $k$ at
the same position $x$), but very weak non-local coupling between
the modes (different positions $x$).
The non-linear dynamical evolution is well developed along the direction
of ${\bf k}$-axis,
rather than ${\bf x}$-axis in the phase space. This quasi-locality has
been justified in weak
nonlinear samples such as the transmitted flux of QSO Ly$\alpha$ absorption
spectrum (Pando, Feng \& Fang 2001). It places the dynamical base of
recovering
initial power spectrum from corresponding weakly evolved field via a
localized mapping in phase space (Feng \& Fang 2000).

This paper is to extend the concept of quasi-locality to fully
nonlinear regime. We try to show that the quasi-locality of the cosmic
clustering in phase space holds not only in the weak nonlinear regime, but
also in nonlinear evolution. Since the nonlinear cosmic density field
can be expressed by the semianalytical halo model (e.g. Cooray \& Sheth
2002, and references therein), our primary interest is to study whether
the quasi-locality could be incorporated in the halo model. We will first
analytically derive the quasi-locality from the halo model, and then
make a numerical test using high resolution $N$-body simulation samples.

The outline of this paper is as follows. \S 2 presents the statistical
criterion of the quasi-local evolution of a density clustering in the
$x$-$k$ phase space.
\S 3 shows that the density field evolution might be spatially
quasi-localized
in the phase space if the cosmic density field can be described by
the halo model. Numerical
tests on these predictions with N-body simulation samples are made
in \S 4. Finally, the conclusions and discussions will be given in \S 5.

\section{Quasi-Locality in $x$-$k$ Space}

\subsection{DWT Variables of the Mass Field}

In physical space (${\bf x}$), the mode is Dirac delta function
$\delta^D({\bf x})$, and cosmic mass density field variable is
$\rho({\bf x})$,
while in scale space (${\bf k}$), the mode is the Fourier bases
$e^{i{\bf k}\cdot{\bf x}}$, and the field variable is
$\hat{\rho}({\bf k})$, which is the Fourier transform of $\rho({\bf x})$.
In hybrid $x$-$k$ phase space, one can use the complete and
orthogonal bases of
the discrete wavelet transform (DWT) as the mode function.
The mass density field is then described by the DWT variables.

Without loss of generality, we introduce the DWT variables by considering a
density field $\rho({\bf x})$ in a cubic box of $0\leq x^i \leq L$,
$i=1,2,3$ and volume $V=L^3$. We first divide the box into cells with volume
$L^3/2^{j_1+j_2+j_3}$, where $j_1, j_2, j_3= 0, 1...$. For a given
${\bf j}\equiv(j_1,j_2,j_3)$, there are $2^{j_1+j_2+j_3}$ cells
labelled by ${\bf l}\equiv (l_1, l_2, l_3)$, and $l_i=0, 1...2^{j_i}-1$.
The cell $(l_1, l_2, l_3)$ occupies the spatial range
$lL/2^{j_i}  < x_i \leq (l+1)L/2^{j_i}$, $i=1,2,3$. Accordingly,
indexes ${\bf j}$ and ${\bf l}$ denote for, respectively, the scale
and the position of the cells. In each dimension, we have
$\Delta x_i= L/2^{j_i}$ and $\Delta k_i = 2\pi / (L/2^j)$, i.e.
$\Delta x_i\Delta k = 2\pi$, or the volume of all cells in the $x$-$k$ space
is $(2\pi)^3$.

Each cell $({\bf j,l})$ supports two compact functions: the scaling
function $\phi_{\bf j,l}({\bf x})$ and the wavelets
$\psi_{\bf j,l}({\bf x})$ (Daubechies 1992,  Fang \& Thews, 1998,
Fang \& Feng, 2000).
Both $\phi_{\bf j,l}({\bf x})$ and $\psi_{\bf j,l}({\bf x})$ are localized
in cell (${\bf j,l}$). The scaling functions $\phi_{\bf j,l}({\bf x})$ are
orthogonormal with respect to index ${\bf l}$ as
\begin{equation}
\int \phi_{\bf j,l}({\bf x})\phi_{\bf j,l'}({\bf x})d{\bf x}
=\delta_{\bf l,l'}.
\end{equation}
The scaling function $\phi_{\bf j,l}({\bf x})$ is a low pass filter
at cell $({\bf j,l})$. The scaling
function coefficient (SFC) of the density field is defined by
\begin{equation}
\epsilon_{\bf j,l}=\int \rho({\bf x})\phi_{\bf j,l}({\bf x})d{\bf x},
\end{equation}
which is proportional to the mean density in cell $({\bf j,l})$.

The wavelets $\psi_{\bf j,l}({\bf x})$ are orthogonormal with
respect to both indexes ${\bf j}$ and ${\bf l}$
\begin{equation}
\int \psi_{\bf j,l}({\bf x})\psi_{\bf j',l'}({\bf x})d{\bf x}
=\delta_{\bf j,j'}\delta_{\bf l,l'}.
\end{equation}
The wavelets $\{\psi_{\bf j,l}({\bf x})\}$ form a complete and orthogonal
base
(mode) in the phase space. Therefore, the density field can be described
by the wavelet function coefficients (WFCs) defined as
\begin{equation}
\tilde{\epsilon}_{\bf j,l}
  =\int d{\bf x}\rho({\bf x})\psi_{\bf j,l}({\bf x}).
\end{equation}
The WFCs $\tilde{\epsilon}_{\bf j,l}$ are the DWT variables of the density
field. The DWT variables
$\tilde{\epsilon}_{\bf j,l}$ is the fluctuation of the density field
around scales ${\bf k}=2\pi {\bf n}/L$, ${\bf n}=(2^{j_1},2^{j_2},2^{j_3})$
located at the cell ${\bf l}$ with size
$\Delta {\bf x}=(L/2^{j_1},L/2^{j_2},L/2^{j_3})$. Since the wavelet
$\psi_{\bf j,l}({\bf x})$ is a band pass filter, in each
dimension, $\tilde{\epsilon}_{\bf j,l}$ is a superposition of
fluctuations filtered in the waveband $k \pm \Delta k/2$, where
$k=2\pi 2^j/L$,
and $\Delta k=2\pi/\Delta x = k$. This decomposition of the
fluctuation is optimized in sense that the size of cell $\Delta x$ is
adaptively chosen to match with the perturbations at a given wavenumber.

The DWT bases generally have vanishing moments, i.e.
\begin{equation}
\int x^t \psi_{\bf j,l}({\bf x})d{\bf x}=0,
\end{equation}
where $t = 0, 1 ... M-1$ and $M \geq 1$. $M$ is dependent on
wavelet. For the wavelet Daubechies $2n$, we have $M=n$. Thus, the
Fourier transform of wavelet $\hat{\psi}_{\bf j,l}({\bf k})=
\int\psi_{\bf j,l}({\bf x})\exp(-i{\bf k}\cdot{\bf x}) d{\bf x}$
has a compact support in the wavenumber space $\{{\bf k}\}$. From
eq.(6) we have $\int \bar{\rho}\psi_{\bf j,l}({\bf x})d{\bf x}=0$
for all ${\bf j,l}$, and thus,
\begin{equation}
\tilde{\epsilon}_{\bf j,l}=\bar{\rho}
\int d{\bf x}\delta({\bf x})\psi_{\bf j,l}({\bf x})
\end{equation}
where the density contrast
$\delta({\bf x})=[\rho({\bf x})-\bar{\rho}]/\bar{\rho}$,
$\bar{\rho}$ is the mean density. Since
$\langle \delta({\bf x})\rangle=0$, we have also
\begin{equation}
\langle \tilde{\epsilon}_{\bf j,l}\rangle=0
\end{equation}

Because the set of wavelets is complete, $\tilde{\epsilon}_{\bf j,l}$
give a complete description of the density field $\rho({\bf x})$, i.e.
one can reconstruct $\rho({\bf x})$ or $\delta({\bf x})$ in terms of variables
$\tilde{\epsilon}_{\bf j,l}$ as
\begin{equation}
\delta({\bf x})=\frac{1}{\bar{\rho}}\sum_{\bf j,l}\tilde{\epsilon}_{\bf j,l}
   \psi_{\bf j,l}({\bf x}).
\end{equation}

\subsection{Quasi-Locality of Gaussian Fields}

The initial density perturbation of the universe $\delta({\bf x}, t_i)$
is believed to be a Gaussian random field with correlation matrix of
the Fourier variables $\hat{\delta}({\bf k}, t_i)$
\begin{equation}
\langle \hat{\delta}({\bf k}, t_i)\hat{\delta}({\bf k'}, t_i)\rangle
  =P({\bf k}, t_i)\delta^K_{\bf k,-k'},
\end{equation}
and all higher order cumulant moments of $\hat{\delta}({\bf k}, t_i)$
vanish. The function $P({\bf k}, t_i)$ of eq.(10) is the initial
power spectrum.
The Kronecker delta function $\delta^K_{\bf k,k'}$ in eq.(10) indicates
that the initial perturbation for each mode ${\bf k}$ is independent, or
localized in $k$-space.

Generally, if the initial Fourier power spectrum, $P({\bf k}, t_i)$, is
colored, i.e. $k$-dependent, the correlation matrix of variables other
than the Fourier mode will no longer be diagonal. For instance, correlation
function in $x$-space will be
$\langle \delta({\bf x}, t_i)\delta({\bf x'}, t_i)\rangle
=\xi({\bf |x-x|}', t_i)$, which is the Fourier counterpart of
$P({\bf k}, t_i)$.
However, the correlation matrix of the DWT variables of a Gaussian field
is always diagonal or quasi-diagonal, regardless the Fourier power
spectrum $P(k)$ is white or colored. That is, the correlation function of
$\tilde{\epsilon}_{\bf j,l}$ is localized with respect to $({\bf j,l})$ as
\begin{equation}
\langle \tilde{\epsilon}_{\bf j,l}(t_i)
      \tilde{\epsilon}_{\bf j',l'}(t_i)\rangle
  \simeq P_{\bf j, l}(t_i) \delta^K_{\bf j,j'}\delta^K_{\bf l,l'},
\end{equation}
where $P_{\bf j,l}$ in eq.(11) is the DWT power spectrum of the field.

The reason of the diagonality of eq.(11) is as follows. First, the WFC
$\tilde{\epsilon}_{\bf j,l}$ is given by a linear superposition of the
Fourier modes $\hat{\delta}({\bf k})$ in the waveband around
$k_i \simeq 2\pi 2^{j_i}/L$ ($i=1,2,3$), $\tilde{\epsilon}_{\bf j,l}$
with different ${\bf j}$ consist of the Fourier modes $\hat{\delta}({\bf k})$
in different ${\bf k}$ bands. While for Gaussian fields, the Fourier
modes in different wave bands are uncorrelated in general [eq.(10)],
and therefore,
there might be no correlation between the DWT modes of ${\bf j}$ and
${\bf j'}$ if ${\bf j\neq j'}$. This yields the quasi-locality of
$\delta^K_{\bf j,j'}$. Second, the phases of the Fourier modes of Gaussian
field are independent and random. For a superposition of the random phased
Fourier modes $\hat{\delta}({\bf k})$ in the band from ${\bf k}$ to
${\bf k+\Delta k }$, the spatial correlation length can not be larger than
that given by the uncertainty relation
$\Delta x \simeq 2\pi/\Delta k \simeq 2^{-j}L$. Moreover, the non-zero
regions of two DWT modes
$\psi_{\bf j,l}$ and $\psi_{\bf j,l'}$
with $({\bf l \neq l'})$ have spatial distance
$\Delta |x_i|\simeq 2^{-j_i}L|l_i-l'_i|$. Consequently, all off-diagonal
elements $({\bf l \neq l'})$ vanish or are much smaller than diagonal
elements, i.e.,
\begin{equation}
  \frac {|\int d{\bf x} \int d{\bf x'}
     \psi_{\bf j,l}({\bf x})\xi( {\bf x- x'}, t_i)
     \psi_{\bf j',l'}({\bf x'})|}
   {\int d{\bf x} \int d{\bf x'}
     \psi_{\bf j,l}({\bf x})\xi( {\bf x- x'}, t_i)
     \psi_{\bf j,l}({\bf x'})
} \ll 1, \hspace{5mm} {\rm if \ \ } {\bf l}\neq {\bf l'}
 \hspace{2mm}
 {\rm or \ \ } {\bf j}\neq {\bf j'}.
\end{equation}
Thus, the correlation function of the DWT variables of a Gaussian
field is rapidly decaying when
$|{\bf l - l'}| \geq 1$ and $|{\bf j - j'}| \geq 1$. The diagonal
correlation function described by eq.(11) is a generic feature of
a Gaussian field in the DWT representation.

It should be pointed out that the WFC correlation function eq.(11)
is different from the ordinary two point correlation function
$\langle \delta({\bf x})\delta({\bf x'})\rangle=\xi({\bf x,x'})$.
The former is the correlation between two modes in $(x-k)$ phase
space, while the later is for two modes in $x$-space. Explicitly,
eq.(11) describes the correlation of perturbation modes in the
waveband $k \rightarrow k+\Delta k$ between positions ${\bf l}$
and ${\bf l'}$, and so, it is sensitive to the phases of modes.
The ordinary two point correlation is not sensitive to the phase
of perturbations. In the DWT analysis, an analogue of the ordinary
two point correlation function is defined by by the correlation
between SFCs, i.e. $\langle\epsilon_{\bf j,l}(t_i)\epsilon_{\bf
j',l'}(t_i)\rangle$. Since the scaling function $\phi_{\bf j,l}$
is a low-pass filter on scale ${\bf j}$ and at position ${\bf l}$,
the correlation function $\langle\epsilon_{\bf
j,l}(t_i)\epsilon_{\bf j',l'}(t_i)\rangle$ behaves in a similar
way as $\langle \delta_R({\bf x})\delta_R({\bf x'})\rangle$, where
$\delta_R({\bf x})$ is a filtered density field smoothed on the
scale $R \sim 2^{-j}L$. However, as $\psi_{\bf j,l}$ is a
high-pass filter, the WFC covariance $\langle
\tilde{\epsilon}_{\bf j,l}\tilde{\epsilon}_{\bf j',l'}\rangle$
shows quite different statistical features from the SFC
correlation, e.g., it is always quasi-diagonal or even fully
diagonal for a Gaussian field. Generally, it has been shown for
many analytically calculable random fields that the SFC
correlation is significantly off-diagonal, while the WFC
correlation is {\it exactly} diagonal (Greiner, Lip \& Carruthers,
1995)

Moreover, within a given volume in the $x$-$k$ space, such as
$Vd^3{\bf k}$, the number of the DWT modes $\{\bf j, l\}$ is the
same as that of the Fourier modes ${\bf k}$. Accordingly, $P({\bf
k}, t_i)$ can be expressed as a linear superposition of $P_{\bf j,
l}(t_i)$, and vice versa.  Equivalently, the Fourier power
spectrum can be replaced by the DWT power spectrum (Fang \& Feng,
2000).

\subsection{Statistical Criterions of Quasi-Locality}

If the evolution of the comic mass field is localized, the evolved
density field $\delta({\bf x})$ at a given spatial point is
determined only by the initial density distribution $\delta({\bf
x}, t_i)$ at the same point. As emphasized in \S 1, this locality
is inconsistent with the non-local behavior of gravitational
clustering. However, the evolution of the comic mass field can be
quasi-localized in sense that the correlation between the DWT
variables of the evolved field is always spatial diagonal if the
initial correlation function is diagonal, such as eq.(11). For
perturbation modes in a waveband $k \rightarrow k+\Delta k$, the
quasi-localized range is $\Delta x\simeq 2\pi/\Delta k$.

A quasi-localized evolution means that the auto-correlation function of
the DWT variables is always diagonal, or quasi-diagonal
$\langle \tilde{\epsilon}_{\bf j,l}\tilde{\epsilon}_{\bf j',l'}\rangle \ll
  \langle \tilde{\epsilon}_{\bf j,l}\tilde{\epsilon}_{\bf j',l}\rangle $
when ${\bf l \neq l'}$, if it is diagonal initially. Thus, one may
place a statistical criterion for the quasi-locality as
\begin{equation}
\kappa_{\bf j,j'}(\Delta {\bf l}) \ll 1,
   \ \ {\rm \ if \ } \Delta {\bf l}\neq 0,
\end{equation}
where
\begin{equation}
\kappa_{\bf j,j'}(\Delta {\bf l})=
\frac{\langle \tilde{\epsilon}_{\bf j,l}\tilde{\epsilon}_{\bf j',l'}\rangle}
{\langle |\tilde{\epsilon}_{\bf j,l}|^2\rangle^{1/2}
\langle|\tilde{\epsilon}_{\bf j',l'}|^2\rangle^{1/2}}.
\end{equation}
This is a normalized correlation function of the DWT modes, i.e.
$\kappa_{\bf j,j}(\Delta {\bf l}=0)=1$.

The auto-correlation function of the DWT variables,
$\langle \tilde{\epsilon}_{\bf j,l}\tilde{\epsilon}_{\bf j',l'}\rangle$,
measures the
correlations between the perturbation modes on scales ${\bf j}=(j_1,j_2,j_3)$
and ${\bf j'}=(j_1',j_2',j_3')$ at two cells with a vector distance given by
$(l_1 L/2^{j_1}-l_1' L/2^{j_1'},l_2 L/2^{j_2}-l_2' L/2^{j_2'},l_3 L/2^{j_3}-l_3' L/2^{j_3'}) $.
In the case of $\Delta l=0$, $\kappa_{\bf j,j'}(0)$ gives the
correlation between fluctuations on scale $j$ and $j'$ at the {\it same}
physical area. Therefore, if condition (13) holds for all redshifts,
the dynamical evolution of the mass field is basically spatial localized
in the DWT bases.
Comparing the condition eq.(13) with eq.(12), we see that the cosmic field
undergoing a local evolution is different from its Gaussian predecessor
by the factor $\delta^K_{\bf j,j'}$, but not $\delta^K_{\bf l,l'}$.
In other words, the evolution leads to the significant scale-scale
$({\bf j,j'})$
coupling, rather than modes at different locations ${\bf l \neq l'}$.

One can also construct the criterions for the quasi-locality using
higher order correlations among the DWT variables
$\tilde{\epsilon}_{\bf j,l}$.
For instance, a $(p+q)$ order statistical criterion is given by
\begin{equation}
C^{p,q}_{\bf j,j'}(\Delta {\bf l} \neq 0) \simeq 1.
\end{equation}
where
\begin{equation}
C^{p,q}_{\bf j,j'}(\Delta {\bf l}) =
\frac{\langle \tilde{\epsilon}^p_{\bf j,l}
\tilde{\epsilon}^q_{\bf j',l'}\rangle}
{\langle \tilde{\epsilon}^{p}_{\bf j,l}\rangle
 \langle \tilde{\epsilon}^{q}_{\bf j',l'}\rangle},
\end{equation}
where $p$ and $q$ can be any even number. Obviously,
$C_{\bf j, j'}^{p,q}\simeq 1$
for Gaussian fields. $C^{p,q}_{\bf j,j'}(0) \neq 1$ corresponds to a local
scale-scale correlation, while $C^{p,q}_{\bf j,j'}(\Delta {\bf l}\neq0)
\neq 1$ a nonlocal scale-scale correlation. The quasi-local evolution of
cosmic mass field requires that nonlocal scale-scale correlation is always
small.

It should be pointed out that the statistical conditions eqs.(13)-(16)
are not trivial because the DWT basis does not subject to the central limit
theorem. If a basis subjects to the central limit theorem, the
corresponding variables will be Gaussian even when the random field is
highly non-Gaussian. In this case, eqs.(13) and (15) may be easily
satisfied, but
it does not imply that the evolution is localized, or quasi-localized.
Statistical measure subjected to the central limit theorem is
unable to capture non-Gaussian features of the evolution.

\section{Quasi-Local Evolution in Halo Model}

We will show, in this section, that the statistical criterions of
\S 2.3 are fulfilled if the cosmic mass field can be described by
the halo model.

\subsection{The Halo Model}

The cosmic clustering is self-similar and hierarchical, as the dynamical
equations of collisionless particles (dark matter) do not have preferred
scales, and admits a self-similar solution as well as the initial density
perturbations are Gaussian and scale free. The halo model further assumes
that all mass in a fully developed cosmic mass
field is bound in halos on various scales (Neyman \& Scott 1952, Scherrer \&
Bertschinger 1991). Thus, the cosmic mass field in non-linear regime is
given by a superposition of the halos
\begin{equation}
\rho({\bf x}) = \sum_i \rho_i({\bf x - x}_i) =
  \sum_i m_i u({\bf x - x_i}, m_i)
= \int d{\bf x'} \sum_i m_i\delta^D({\bf x'-x_i})u({\bf x - x'}, m_i),
 \end{equation}
where $\rho_i({\bf x - x}_i)$ is the density profiles of halo with mass
$m_i$ at position ${\bf x_i}$, and
$u({\bf x - x_i}, m_i)$ is the density profile normalized by
$\int d{\bf x} \rho_i({\bf x - x}_i)=\int d{\bf x}m_i u({\bf x - x}_i, m_i)
=m_i$.

There are several different versions of the halo density profiles,
such as $u({\bf x}, m_i)\propto
1/(r/r_i)^{\alpha}[1+(r/r_i)]^{\beta}$ with $\alpha=1, \beta=2$
(Navarro, Frenk \& White 1996), and $u({\bf x}, m_i)\propto
1/(r/r_i)^{\alpha}[1+(r/r_i)^{\beta}]$ with $\alpha=3/2,
\beta=3/2$ (Moore et al. 1999). A common feature of the halo
density profiles is self-similar, which implies that the indexes
$\alpha$ and $\beta$ should be mass-independent. The mass
dependence is only given by $r_i$, which characterizes the size of
the $m_i$ halo. The details of the profiles are indifferent for
the problem we try to study below. What is important for us is
only that one can set a self-similar upper limit to the normalized
halo density profile as
\begin{equation}
u({\bf x}, m_i) < C(r/r_i)^{-\gamma}.
\end{equation}
where $r=|{\bf x}|$, $C$ is a constant and the index $\gamma$ is
mass-independent. The $m_i$-dependence of $r_i$ are not stronger than
a power law as $r_i \propto m_i^{\eta}$.

The halo model also assumes that the halo-halo correlation
function on scales larger than the size of halos is given by the
two-point correlation functions of the linear Gaussian field with
a linear bias correction, or by the correlation functions of
quasi-linear field. Therefore, no assumption about higher order
halo-halo correlations on large scales is needed.

In this model, the time-dependence of the field is mainly given by the
mass function of halos, $n(m, t)$, which is the number density of the
halos with mass $m$ at time $t$. In the Press \& Schechter formalism
(1974), the mass function is determined by the power spectrum of the
initial Gaussian density perturbation. Moreover, the self-similarity
of the halo density profiles insure that eq.(18) holds for all time.
The cosmic evolution only leads to the parameters on the r.h.s of eq.(18)
to be time-dependent.

\subsection{Quasi-Locality of the DWT Correlation Function}

With eq.(17), the DWT variable of the cosmic mass field in the halo model
is given by
\begin{equation}
\tilde{\epsilon}_{\bf j,l}
  =\int d{\bf x}\rho({\bf x})\psi_{\bf j,l}({\bf x})
  = \sum_i m_i \int d{\bf x}u({\bf x - x_i}, m_i)\psi_{\bf j,l}({\bf x}).
\end{equation}
The auto-correlation function of the DWT variables is then
\begin{equation}
\langle \tilde{\epsilon}_{\bf j,l}\tilde{\epsilon}_{\bf j',l'}\rangle
    =\langle \tilde{\epsilon}_{\bf j,l}\tilde{\epsilon}_{\bf j',l'}\rangle^h
  + \langle \tilde{\epsilon}_{\bf j,l}
  \tilde{\epsilon}_{\bf j',l'}\rangle^{hh},
\end{equation}
where the first and second terms on the r.h.s. are usually called,
respectively,
1- and 2-halo terms. They can be written in the explicit form,
\begin{eqnarray}
\langle \tilde{\epsilon}_{\bf j,l}\tilde{\epsilon}_{\bf j',l'}\rangle^h
  & = & \int  dm n(m)m^2
   \int d{\bf x}_1u({\bf x}_1, m)
\\ \nonumber
  & &  \int d{\bf x} \int d{\bf x'}
     \psi_{\bf j,l}({\bf x})u({\bf x_1+ x- x'}, m)
\psi_{\bf j',l'}({\bf x'})
\end{eqnarray}
\begin{eqnarray}
\langle \tilde{\epsilon}_{\bf j,l}\tilde{\epsilon}_{\bf j',l'}\rangle^{hh}
& = & \int  dm_1 n(m_1)m_1\int  dm_2 n(m_2)m_2
 \int d{\bf x}_1\int d{\bf x}_2 u({\bf x_1}, m_1) u({\bf x_2}, m_2)
 \\ \nonumber
 & & \int d{\bf x} \int d{\bf x'}
     \psi_{\bf j,l}({\bf x})\xi({\bf x - x' + x_1-x_2}, m_1, m_2)
     \psi_{\bf j',l'}({\bf x'}),
\end{eqnarray}
where
$n(m)=\langle \sum_i\delta^D(m-m_i)\delta^D({\bf x - x_i}) \rangle$ is
the number density of halos with mass $m$, and
$\xi({\bf x - x'}, m_1, m_2)
  =\langle\sum_{i\neq j}\delta^D(m_1-m_i)\delta^D({\bf x - x_i})
   \delta^D(m_2-m_i)\delta^D({\bf x' - x_i})\rangle/n(m_1)n(m_2)$ is the
halo-halo correlation function.

We show now that the DWT correlation function
$\langle \tilde{\epsilon}_{\bf j,l}\tilde{\epsilon}_{\bf j',l'}\rangle$
is quasi-diagonal, or fast decaying with respect to the spatial distance
${\bf l-l'}$.
First, consider the 2-halo terms eq. (22). According to the halo model,
the two-point correlation function
$\xi({\bf x_1 - x_2}, m_1, m_2)$ is determined by the linearly Gaussian
density field. In fact, as having been discussed in \S 2.2, the DWT
correlation
function of a Gaussian field is generally diagonal [eq.(12)], regardless
the Fourier power spectrum is white or colored. Therefore,
the DWT correlation function
$\langle \tilde{\epsilon}_{\bf j,l}\tilde{\epsilon}_{\bf j',l'}\rangle$
contributed from the 2-halo term should be quasi-local.

The DWT integral in the 2-halo term of eqs.(22) is not completely the
same as eq.(12), as the halo-halo correlation function in eq.(22) is
$\xi({\bf x - x' + x_1-x_2})$, while in eq.(12) is $\xi({\bf x - x'})$. For
halos with size less than the scale ${\bf j}$ considered, the factor
$|{\bf x_1-x_2}|$ in correlation function $\xi^{hh}$ is smaller than
the size of the DWT mode, $L/2^{j}$, and so, the factor
${\bf x_1-x_2}$ can be ignored in comparison with the variable ${\bf x-x'}$.
Moreover, by definition of the halo model, the correlation
functions $\xi^{hh}$ does not include the contributions from halos with
sizes larger than $|{\bf x-x'}|$. Thus, the cases of
$|{\bf x_1-x_2}|>|{\bf x - x'}|$ can always be ignored. The 2-halo term
essentially follows eq.(12), and is always approximately diagonal with
respect to spatial indexes ${\bf l}$ and ${\bf l'}$.

To analyze the 1-halo term eq.(21), we use the following theorem of
wavelets (Tewfik \& Kim, 1992). Because the DWT bases is self-similar,
for any 1-D power law function $f(x) \propto x^{-\gamma}$, we have
\begin{equation}
\left |\int dx \int d x' \psi_{j,l}(x)f(x - x')\psi_{j,l'}( x') \right |
     < C'2^{-j}|l -l'|^{-\gamma -2M},
\end{equation}
where $\psi_{j,l}(x)$ is wavelet in 1-D space, and $C'$ is a constant.
Therefore, while using wavelets with large enough $M$ [eq.(5)],
the integral eq.(23) is quickly decaying with the spatial distance $|l -l'|$.
In other words, besides two nearest position of $l -l' = \pm 1$, there
is no correlation between modes at different position $l \neq l'$.

The 3-D integral $I(|{\bf l-l}'|) \equiv |\int d{\bf x} \int d{\bf x'}
     \psi_{\bf j,l}({\bf x})u({\bf x_1+ x- x'}, m)
\psi_{\bf j,l'}({\bf x'})|$ in eq.(21) has the similar structure as
eq.(23). Using the upper limit eq.(18), one can estimate the decaying
of the integral $I(|{\bf l-l}'|)$ with $|{\bf l-l'}|$.
When  $|{\bf x}_1|< |{\bf x- x'}|$, we can expand the function
$|{\bf x_1+ x- x'}|^{-\gamma}$ in terms of $|{\bf x}_1|/|{\bf x- x'}|$.
Thus, applying eq.(23) term by term, we have the decaying behavior of $I$ at
least
as fast as $2^{-j_1+j_2+j_3}|{\bf l-l'}|^{-\gamma-2M}$. Similarly,
in the case of
$|{\bf x}_1|> |{\bf x - x'}|$, one can expand the function
$|{\bf x_1+ x- x'}|^{-\gamma}$ by the factor $|{\bf x- x'}|/|{\bf x}_1|$.
Following
from eq.(23), the decaying behavior of $I$ is then
decreasing with $|{\bf l-l'}|$ as
$2^{-j_1+j_2+j_3}|{\bf l-l'}|^{-4M}$. Accordingly, the 1-halo
term in the correlation between modes ${\bf j, l}$ and ${\bf j,l'}$ will
generally decay as $|{\bf l-l'}|^{-\gamma-2M}$ or
$|{\bf l-l'}|^{-4M}$. The correlation function
$\langle \tilde{\epsilon}_{\bf j,l}\tilde{\epsilon}_{\bf j,l'}\rangle^h$
is then approximately diagonal w.r.t. the spatial index ${\bf l}$ and
${\bf l'}$. This result is largely valid due to $r_i$ varying with
$m_i$ by a power law.

Proceeding in the similar way as above, we can also show the
diagonality of the correlation function
$\langle \tilde{\epsilon}_{\bf j,l}\tilde{\epsilon}_{\bf j',l'}\rangle^h$.
In this case, instead of eq.(23), we use the following theorem
(Tewfik \& Kim, 1992)
\begin{equation}
\left |\int dx \int d x' \psi_{j,l}(x)f(x - x')\psi_{j',l'}( x') \right |
     < C'2^{-j}|l -(l'/2^{j'-j})|^{-\gamma -2M},  \hspace{2mm} {\rm if}
  \ j'>j.
\end{equation}
Consider that the cell $(j, l)$ has the same physical position as cell
$(j',l')$ if $l'=l\times 2^{j'-j}$, the theorem eq.(24) also yields
that the 1-halo term of the two modes ${\bf j,l}$ and ${\bf j',l'}$ will
decline with the physical distance between the two modes as
$|{\bf l-(l'}/2^{j'-j})|^{-\gamma-2M}$ or $|{\bf l-(l'}/2^{j'-j})|^{-4M}$ .

Based on above discussions, one may draw the conclusion that if the
halo model is a good approximation
to the cosmic density field $\rho({\bf x})$ in the nonlinear regime at all
time, their correlation matrix in the DWT representation remains
quasi-diagonal
forever, and the evolution is quasi-local. This result is based on the
self-similarity
of the density profiles of halos and the weakly nonlinear correlation
between halos.
The self-similar scaling ensures that
the non-local correlation among the DWT variable is uniformly suppressed,
independent of the mass of halos. Mathematically, the
correlations between the DWT modes of perturbations at different physical
places
are uniformly converging to zero with the increasing of $M$ if the index
$\gamma$ is mass-independent.

\subsection{Quasi-Locality of Higher Order Statistics}

To show the quasi-locality of higher order statistical criterion, we use the
hierarchical clustering or linked-pair relation (Peebles 1980), which
is found to be consistent with the halo model. For the 3$^{rd}$ order
correlations, the linked-pair relation is
\begin{eqnarray}
\langle\tilde{\epsilon}_{\bf j'',l^{''}}
\tilde{\epsilon}_{\bf j_3,l_3} \rangle \langle\tilde{\epsilon}_{\bf j'',l^{''}}
\tilde{\epsilon}_{\bf j_3,l_3} \rangle
 & \simeq & Q_3
[\langle \delta({\bf x^1})\delta({\bf x^2})\rangle \langle
\delta({\bf x^1})\delta({\bf x^3}) \rangle  \\ \nonumber
 & & + {\rm \ 2 \ terms \ with \ cyc. \ permutations}],
\end{eqnarray}
where the coefficient $Q_3$ might be scale-dependent. Subjecting eq.(25)
to a DWT by 3$^{rd}$ order basis
$\psi_{\bf j_1,l_1}({\bf x_1})\psi_{\bf j_2,l_2}({\bf x_2})
\psi_{\bf j_3,l_3}({\bf x_3})$, we have
\begin{eqnarray}
\langle \tilde{\epsilon}_{\bf j_1,l_1}
        \tilde{\epsilon}_{\bf j_2,l_2}
        \tilde{\epsilon}_{\bf j_3,l_3} \rangle  & \simeq &
Q_3 \sum_{\bf j', l'; j'', l''}a^3_{\bf j_1, l_1; j', l'; j'', l^{''}}
  [\langle \tilde{\epsilon}_{\bf j',l^{'}}
\tilde{\epsilon}_{\bf j_2,l_2}\rangle
\langle\tilde{\epsilon}_{\bf j'',l^{''}}
\tilde{\epsilon}_{\bf j_3,l_3} \rangle  \\ \nonumber
 & & + {\rm \ 2 \ terms \ with \ cyc. \ permutations}].
\end{eqnarray}
where $a^3_{\bf l^1, l^{'1},l^{''1}}$ is given by the 3-wavelet integral,
\begin{equation}
a^3_{\bf j_1, l_1; j', l'; j'', l^{''}}
=\int \psi_{\bf j_1,l_1}({\bf x})\psi_{\bf j',l^{'}}({\bf x})
  \psi_{\bf j'',l^{''}}({\bf x}) d{\bf x}.
\end{equation}
Since $\psi_{\bf j,l}({\bf x})$ is localized in the cell (${\bf j,l}$),
$a^3$ is
significant only if the three cells (${\bf j_1, l_1}$),
(${\bf j', l'}$) and (${\bf j'', l^{''}}$) coincide with each other at
the same physical area. Thus, by virtue of the locality of correlations
$\langle \tilde{\epsilon}_{\bf j',l^{'}}\tilde{\epsilon}_{\bf j_2,l_2}\rangle$
and
$\langle\tilde{\epsilon}_{\bf j'',l^{''}}
\tilde{\epsilon}_{\bf j_3,l_3} \rangle$ (\S 3.2), it is easy to see that
$\langle \tilde{\epsilon}_{\bf j_1,l_1}\tilde{\epsilon}_{\bf j_2,l_2}
        \tilde{\epsilon}_{\bf j_3,l_3} \rangle $
is small if the cells (${\bf j_1, l_1}$), (${\bf j_2, l_2}$)
and (${\bf j_3, l_3}$) are disjoint in the physical space. Since $Q_3$ does
not depend on ${\bf x}$, the result of locality will keep valid when $Q_3$ is
scale-dependent,

Obviously, the 3$^{nd}$ order result can be generalized to $n^{th}$ order
DWT correlation function. The integral of $n$ wavelets $\psi_{\bf j_1, l_1}$,
$\psi_{\bf j_2, l_2}$,...$\psi_{\bf j_n, l_n}$ is zero or very small otherwise
the $n$ cells (${\bf j_1, l_1}$), (${\bf j_2, l_2}$)...
(${\bf j_n, l_n}$) coincide in the same physical area. In addition, all
terms on the r.h.s. of the hierarchical clustering relation consist of
linked 2$^{nd}$ DWT correlation function, the $n^{th}$ order
DWT correlation function
$\langle \tilde{\epsilon}_{\bf j_1,l_1}\tilde{\epsilon}_{\bf j_2,l_2}
      ... \tilde{\epsilon}_{\bf j_n,l_n} \rangle$ should be localized.
Thus, the criterion eq.(15) and other higher order criterions will
be satisfied in general.

In summary, if the cosmic density field is evolved self-similarly
 from an initially Gaussian field, the spatial quasi-locality is true at all
  time, i.e.
\begin{itemize}
\item the second and higher order correlation functions of the DWT variables,
$\tilde{\epsilon}_{\bf j,l}$ of the evolved field is quasi-diagonal
with respect to the position index ${\bf l}$.
\end{itemize}
For those type of fields, the possible non-Gaussian features with the
DWT variables are mainly
\begin{itemize}
\item non-Gaussian one-point distribution of the DWT variables
 $\tilde{\epsilon}_{\bf j,l}$;
\item local scale-scale correlation among the DWT variables.
\end{itemize}
Above three points are the major theoretical results of this paper.

\section{Testing with N-body Simulation Samples}

\subsection{Samples}

To demonstrate the quasi-locality of the evolved cosmic density field, we use
samples produced by Jing \& Suto (2002). The samples are given by high
resolution N-body simulation, running with the vectorized-parallelized
P$^3$M code. The cosmological model was taken to be LCDM model specified
by parameters
$(\Omega_0,\Lambda,\sigma_8,\Gamma)=(0.3,0.7,0.9,0.2)$.
The primordial density fluctuation is assumed to obey the
Gaussian statistics (this is important for us), and the power spectrum is
of the Harrison-Zel'dovich type. The linear transfer function for
the dark matter power spectrum is taken from Bardeen et al (1986).

The simulation was performed in a periodic, cubical box of size
100h$^{-1}$Mpc with a 1200$^3$ grid points for the Particle-Mesh (PM) force
computation, and 512$^3$ particles. The short-range force is compensated
for the PM force calculation at the separation less than $\epsilon=2.7H$,
where $H$ is the mesh cell size. The simulations are evolved by 1200 time
steps from the initial redshift $z_i=72$. The force resolution is
20 h$^{-1}$ kpc for the linear density softening form. It is noted
that our statistic
tests are performed on scales $\geq 0.2$ h$^{-1}$ Mpc.

We have one realization. In the practical computations, we divide the
100 h$^{-1}$ Mpc simulation box into 8 subboxes each with size
$L=$ 50h$^{-1}$ Mpc.
Accordingly, the ensemble average and 1$\sigma$ variance are obtained
from those 8 subboxes.

\subsection{Two-Point Correlation Functions}

Before showing the quasi-locality, we first calculate the correlation
function of the SFCs, i.e.
$\langle \epsilon_{\bf j,l} \epsilon_{\bf j,l'}\rangle$. As has been
discussed  in \S 2.2, the correlation function
$\langle \epsilon_{\bf j,l} \epsilon_{\bf j,l'}\rangle$ is actually
analogue to the ordinary two-point correlation function
$\xi({\bf r})$, where ${\bf r}={\bf x-x'}$. Since $\epsilon_{\bf j,l}$
is a filtered density field smoothed by the scaling function
on the scale ${\bf j}$, it is expected that the correlation
function $\langle \epsilon_{\bf j,l} \epsilon_{\bf j,l'}\rangle$
will display similar feature as $\xi({\bf r})$.

Figure 1 presents the $r$-dependence of
$\langle \epsilon_{\bf j,l} \epsilon_{\bf j,l'}\rangle$
at ${\bf j}=(7, 7, 7)$, corresponding to the smoothed field filtered
on the linear
scale $50/2^7=0.39$ h$^{-1}$ Mpc. The spatial distance
between the cells ${\bf l}$ and ${\bf l'}$ is
$r=|{\bf l-l'}|50/2^7$ h$^{-1}$ Mpc. In this calculation, we applied
wavelet
Daubechies 4 (Daubechies, 1992). As expected, the $r$- or
$|{\bf l-l'}|$-dependence of the correlation function
$\langle \epsilon_{\bf j,l} \epsilon_{\bf j,l'}\rangle$ shows the
standard power law,
$\langle \epsilon_{\bf j,l} \epsilon_{\bf j,l'}\rangle \propto
r^{-\alpha}$ with the index $\alpha \simeq 2$.
That is, this correlation function is not localized. Meanwhile in
Fig.1, the contributions of the 1-halo and 2-halo
terms are also plotted, respectively. The 1-halo term is calculated
from eq.(21) with the NFW density profiles of halos, and the 2-halo
term is given by eq.(22) in which the correlation function
$\xi({\bf r})$ is given by the linear power spectrum. The nonlinear
clustering is largely due to the 1-halo term. Therefore, one can see
that the nonlinear evolution of gravitational clustering cause correlations
on scales of several Mpc.

\subsection{Justifying the Quasi-Locality}

We now study the quasi-locality of the clustering with the correlation
function
$\langle \tilde{\epsilon}_{\bf j,l}\tilde{\epsilon}_{\bf j',l'}\rangle$,
which is used in the criterion $\kappa_{\bf j,j'}(\Delta {\bf l})$
[eq.(13)]. First, we take the same parameter as Fig. 1,
${\bf j = j'}=(7,7,7)$,
and $r=|\Delta{\bf l}|50/2^j=|{\bf l-l'}|50/2^j$ h$^{-1}$ Mpc, and also
we used wavelet Daubechies 4 (Daubechies, 1992), which has
$M=2$. The result of DWT correlation function
$\langle \tilde{\epsilon}_{\bf j,l}\tilde{\epsilon}_{\bf j,l'}\rangle$.
 is shown in Fig. 2. The solid
circle at $r=0$ in Fig. 2 corresponds to $\Delta {\bf l}=0$, or
${\bf l =l'}$, and other solid circles from left to right correspond,
successively, to $\Delta {\bf l}=1$, 2, 3...

 From Fig. 2, we can see immediately that the shape of the $r$- or
$|\Delta{\bf l}|$-dependence of
$\langle \tilde{\epsilon}_{\bf j,l}\tilde{\epsilon}_{\bf j,l'}\rangle$
is quite
different from the ``standard'' power law. The correlation function
$\langle \tilde{\epsilon}_{\bf j,l}\tilde{\epsilon}_{\bf j,l'}\rangle$
is non-zero mainly at point $r=0$ or
$|\Delta{\bf l}|=0$. At $|\Delta{\bf l}|=1$, the correlation
function
$\langle \tilde{\epsilon}_{\bf j,l}\tilde{\epsilon}_{\bf j,l\pm 1}\rangle$
drops to tiny values around $\sim 0$. For $|\Delta{\bf l}| >1$, the
correlation function basically is zero.
The correlation length in terms of
the position index ${\bf l}$ is approximately zero, namely, the covariance
$\langle \tilde{\epsilon}_{\bf j,l}\tilde{\epsilon}_{\bf j,l'}\rangle$
is diagonal. This is the spatial quasi-locality.
In Fig. 2, we also plot the 1- and 2-halo terms,
$\langle \tilde{\epsilon}_{\bf j,l}\tilde{\epsilon}_{\bf j,l'}\rangle^h$
and
$\langle \tilde{\epsilon}_{\bf j,l}\tilde{\epsilon}_{\bf j,l'}\rangle^{hh}$.
Although 1-halo term is dominated by massive halos, the covariance
$\langle \tilde{\epsilon}_{\bf j,l}\tilde{\epsilon}_{\bf j,l'}\rangle^h$
is also perfectly quasi-localized because of the self-similarity of density
profiles of massive halos. The 2-halo term is zero at $r=0$, because,
by definition, eq.(22) does not contain the contribution of
autocorrelations of halos.

Figure 3 presents $\kappa_{\bf j,j'}(\Delta{\bf l})$ vs. $r$,
for ${\bf j}={\bf j'}=(4, 4, 4)$, $(5,5,5)$ and $(6,6,6)$. The
physical distance $r$ is the same as Fig. 2, given by
$r=|\Delta {\bf l}|50/2^j=|{\bf l-l'}|50/2^j$ h$^{-1}$ Mpc. The
solid circle at $r=0$ corresponds to $|\Delta {\bf l}|=0$, at which, by
definition, we have the normalization $\kappa_{\bf j,j}(0)=1$. Other
points from small to
large values of $r$ correspond to $\Delta {\bf l}=$ 1,2,.... successively.
Clearly, all $\kappa_{\bf j,j}(\Delta{\bf l})$ for $\Delta l > 0$ are less
than $10^{-6}$, which is actually from the noises of sample. The result
implies that
for all calculable points of $r\geq 0$, the
correlation is negligible, and satisfies the criterion eq.(13).

We also calculated $\kappa_{\bf j,j'}(\Delta{\bf l})$ for modes of
${\bf j=j'}=(j_1,j_2,j_3)$, but $j_1 \neq j_2 \neq j_3$.
Most of these cases shows
$\kappa_{\bf j,j'}(\Delta{\bf l})\simeq 0$ if $|\Delta{\bf l}|>0$. Only
exception is for the cases of $j_1, j_2 < j_3$, and
$|\Delta {\bf l}|= |l_3-l'_3|=1$.
As an example, Fig. 4 presents $\kappa_{\bf j,j}(\Delta {\bf l})$
for modes ${\bf j}=(5,5,6)$, and $r= |{\bf l_3-l'_3}|50/2^{j_3}$. It
shows that
$\langle \tilde{\epsilon}_{\bf j,l}\tilde{\epsilon}_{\bf j,l'}\rangle=0$
if $|l_3-l'_3| > 1$. The non-zero value at $|l_3-l'_3| =1$ is about
20\% of that
at $\Delta l =0$. This result is consistent with the theorem (23),
which only requires the suppression for modes $|l-l'| > 1$, but may not work
for $|l-l'| = 1$. It implies that the clustering may give rise to the
correlations between
nearest neighbor cells in phase space. However, the cell
resolved by $j_1, j_2 < j_3$ is a rectangle in the physical space, and
the shortest
edge is given by $j_3$, the physics distance of $|l_3-l'_3|=1$ is still less
than whole size of the rectangle. Thus, it could be concluded that
the covariance
$\langle \tilde{\epsilon}_{\bf j,l}\tilde{\epsilon}_{\bf j,l'}\rangle$
is always quasi-diagonal in the sense that all members with ${\bf l}$ and
${\bf l'}$ are almost zero if the distance $r$ between ${\bf l}$ and
${\bf l'}$ is larger than the size of the cell $(j_1,j_2,j_3)$ considered.
In Fig. 4, we show also a result calculated
with wavelets Daubechies 6 (D6), for which $M=3$. It yields about the
same results as D4.

For the correlation between modes $({\bf j,l})$ and
$({\bf j',l'})$ with ${\bf j\neq j'}$, we use the criterion
$\kappa_{\bf j,j'}$ of eq.(14). In this case, the physical
distance between two cells is
$r=|{\bf r}|$ and $r_i=|l_i/2^{j_i} -l'_i/2^{j'_i}|L$ h$^{-1}$ Mpc.
Figure 5 plots $\kappa_{\bf j,j'}(\Delta {\bf l})$ vs. $r$ for modes
${\bf j}=(j,j,j)$, ${\bf j'}=(j+1,j+1,j+1)$, and $j=3, 4, 5$ and 6.
All the values of $\kappa_{\bf j,j'}(\Delta {\bf l})$ in Fig. 5 are not
larger than $10^{-5}$, and are much less than the diagonal terms
$\langle \tilde{\epsilon}^2_{\bf j,l}\rangle$, or
$\langle \tilde{\epsilon}^2_{\bf j,l'}\rangle$. We found this result
is generally true for all the cases of  ${\bf j \neq j'}$.
That is, the second order correlation between two modes with different
scales ${\bf j}$ and ${\bf j'}$ are always negligible, regardless the
indices ${\bf l}$ and ${\bf l'}$. In other words, the covariance of the
WFC variables
$\langle \tilde{\epsilon}_{\bf j,l}\tilde{\epsilon}_{\bf j',l'}\rangle$
generally are quasi-diagonal.

As for the high order statistics by criterion eqs.(15) and (16), we
can cite some previous calculations of the non-local scale-scale
correlation defined by
\begin{equation}
C^{2,2}_{\bf j,j}(\Delta {\bf l}) =
\frac{\langle \tilde{\epsilon}^2_{\bf j,l}
  \tilde{\epsilon}^2_{\bf j,l+\Delta l}\rangle}
{\langle \tilde{\epsilon}^{2}_{\bf j,l}\rangle
 \langle \tilde{\epsilon}^{2}_{\bf j,l+\Delta l}\rangle}.
\end{equation}
which is the criterion eq.(15) with $p=q=2$. It has been shown that
either for the APM bright galaxy catalog (Loveday et al. 1992) or mock
samples of galaxy survey (Cole et al. 1998), the non-local
scale-scale correlation always yields
$|C^{2,2}_{\bf j,j}(\Delta {\bf l})-1|\ll 1$ if $\Delta {\bf l}>0$
 (Feng, Deng \& Fang 2000). Although this work was not for
addressing the problem of the quasi-locality, the result did support
the quasi-locality up to the 4$^{th}$ order statistics.

\subsection{Non-Gaussianity Revealed by DWT Variables}

As discussed in \S 2.3, it is necessary to show that the random variables
$\tilde{\epsilon}_{\bf j,l}$ are
non-Gaussian for evolved fields. The possible non-Gaussian features with
the DWT variables are (1) non-Gaussian one-point distribution of
$\tilde{\epsilon}_{\bf j,l}$, and (2) local scale-scale correlations (\S 3.3).

In Fig. 6, we plot the one-point distribution on scales ${\bf j}=(j,j,j,)$
and $j=$ 5, 6, 7 and 8. It illustrates that the kurtosis of the one-point
distribution is high. The PDF (probability distribution function) is
approximately
lognormal. The 4$^{th}$ order local scale-scale correlation
$C^{2,2}_{\bf j,j}(\Delta {\bf l}=0)$ is plotted in Fig. 7. It shows
$C^{2,2}_{\bf j,j}(\Delta {\bf l}=0) \gg 1$ on small scales
($j=$ 5, \ 6 and 7),
while random data gives $C^{2,2}_{\bf j,j}(\Delta {\bf l}=0) = 1$ on
all scales. The evolved field is highly non-Gaussian, although it is always
quasi-localized.

\section{Discussions and Conclusions}

We showed that the cosmic clustering behavior is quasi-localized. If the
field is viewed by the DWT modes in phase space, the nonlinear evolution
will give rise to the coupling between modes on different scales but
in the same physical area, and the coupling between modes at different
position is weak. The quasi-local evolution means that, if the initial
perturbations in a waveband $k \pm \Delta k/2$ and at different space range
$\Delta x$ is uncorrelated, the evolved perturbations in this waveband at
different space range $\Delta x$ will also be uncorrelated, or very weakly
correlated. In this sense,
the nonlinear evolution has memory of its initial spatial correlation in
the phase space. This memory is essentially from the hierarchical and
self-similar feature of the mass field evolution. The density profiles of
massive halos obey the scaling law [eq.(18)], and therefore, the contributions
to the non-local correlation function from various halos are uniformly
suppressed.

It has been realized about ten years ago that some random fields generated
by a self-similar hierarchical process generally shows locality of their
auto-correlation function in the phase space, if the initial field is local,
like a Gaussian field (Ramanathan \& Zeitouni, 1991; Tewfik \&  Kim 1992;
Flandrin, 1992). Later, this result are found to be correct for
various models of structure formations via hierarchical cascade stochastic
processes (Greiner et al. 1996, Greiner, Eggers \& Lipa, 1998). These studies
implies that the local evolution and initial perturbation memory seems to
be generic
of self-similar hierarchical fields, regardless the details of the
hierarchical process. It has been pointed out that models for realizing the
self-similar hierarchical evolution of cosmic mass field, such as the
fractal hierarchy clustering model (Soneira \& Peebles
1977), the block model (Cole \& Kaiser 1988), merging cell (Rodrigues \&
Thomas 1996), have the same mathematical structures as hierarchical
cascade stochastic models applied in other fields (Pando et al 1998, Feng,
Pando \& Fang 2001). Obviously, the local evolution
can be straightforward obtained in those models.

The DWT analysis is effective to reveal the quasi-locality in
phase space. Such quasi-locality is hardly described by the
Fourier modes $\hat{\delta}({\bf k})$, as the information of
spatial positions is stored in the phases of all Fourier modes.
Moreover, the Fourier amplitudes $|\hat{\delta}({\bf k})|$ subject
to the central limit theorem, and are insensitive to
non-Gaussianity. The wavelet basis, however does not subject to
the central limit theorem (Pando \& Fang 1998), which enable us to
measure all the quasi-local features with the statistics of
$\tilde{\epsilon}_{\bf j,l}$.

The quasi-locality of the DWT correlation is essential for recovery of
the primordial power spectrum using a localized mapping in phase space. Such
mapping has been developed in recovering the initial Gaussian power
spectrum from evolved field in the quasi-linear regime (Feng \& Fang 2000).
By virtue of the quasi-locality in fully developed fields, we would be able
to generalize the method of localized mapping in phase space to highly
non-linear regime.

The quasi-local evolution may also provide the dynamical base for the
lognormal model (Bi, 1993; Bi \& Davidsen 1997, Jones
1999). The basic assumption of the lognormal model is that the non-linear
field
can be approximately found from the corresponding linear Gaussian field
by a local exponential mapping. The local mapping is supported by the
quasi-local evolution. We see from Fig. 6 that the PDF of evolved field
is about lognormal. Therefore, in the context of quasi-local evolution, a
local (exponential) mapping from the linear Gaussian field to a lognormal
field might be a reasonable sketch of the nonlinear evolution of
the cosmic density field.

\acknowledgments

We thank Dr. Y.P. Jing for kindly providing his N-body simulation data.
LLF acknowledges supports from the National Science Foundation of China
(NSFC) and National Key Basic Research Science Foundation.

\clearpage

\begin{figure}
\figurenum{1}\epsscale{0.85}\plotone{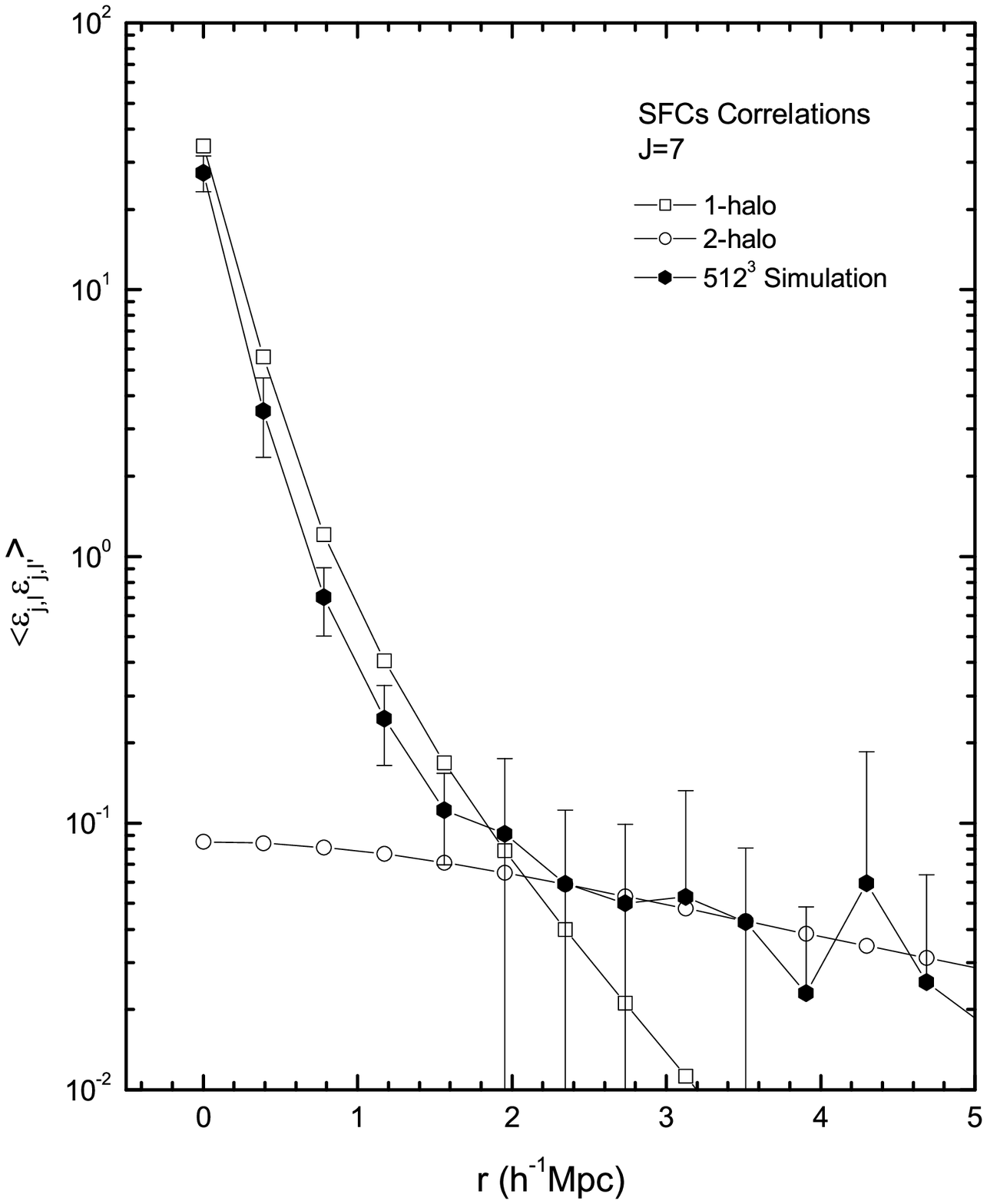}\caption{The SFC
correlation function $\langle\epsilon_{\bf j,l}\epsilon_{\bf
j,l'}\rangle$ vs. $r$ for the simulation data (hexagon), 1-halo
(square) and 2-halo (circle) terms. The scale ${\bf j}$ is taken
to be $(7,7,7)$. The physical distance is given by $r=|{\bf
l-l'}|50/2^j$ h$^{-1}$ Mpc. }
\end{figure}

\begin{figure}
\figurenum{2}\epsscale{0.85}\plotone{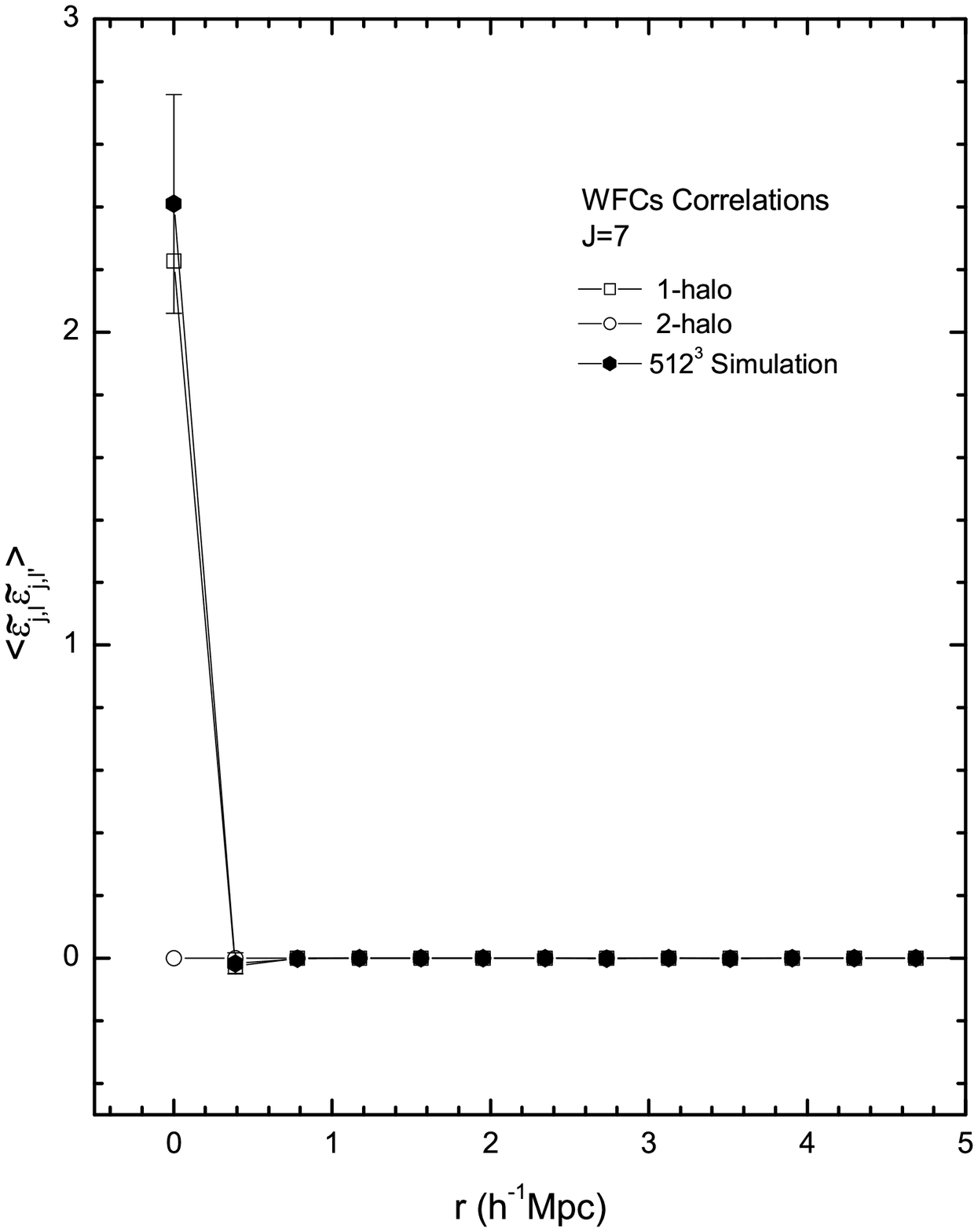}\caption{The DWT
variable (WFC) correlation function $\langle\tilde{\epsilon}_{\bf
j,l}\tilde{\epsilon}_{\bf j,l'}\rangle$ vs. $r$ for the simulation
data (hexagon), 1-halo (square) and 2-halo (circle) terms. The
scale ${\bf j}$ is taken to be $(7,7,7)$. The physical distance is
given by $r=|{\bf l-l'}|50/2^j$ h$^{-1}$ Mpc. }
\end{figure}

\begin{figure}
\figurenum{3}\epsscale{0.85}\plotone{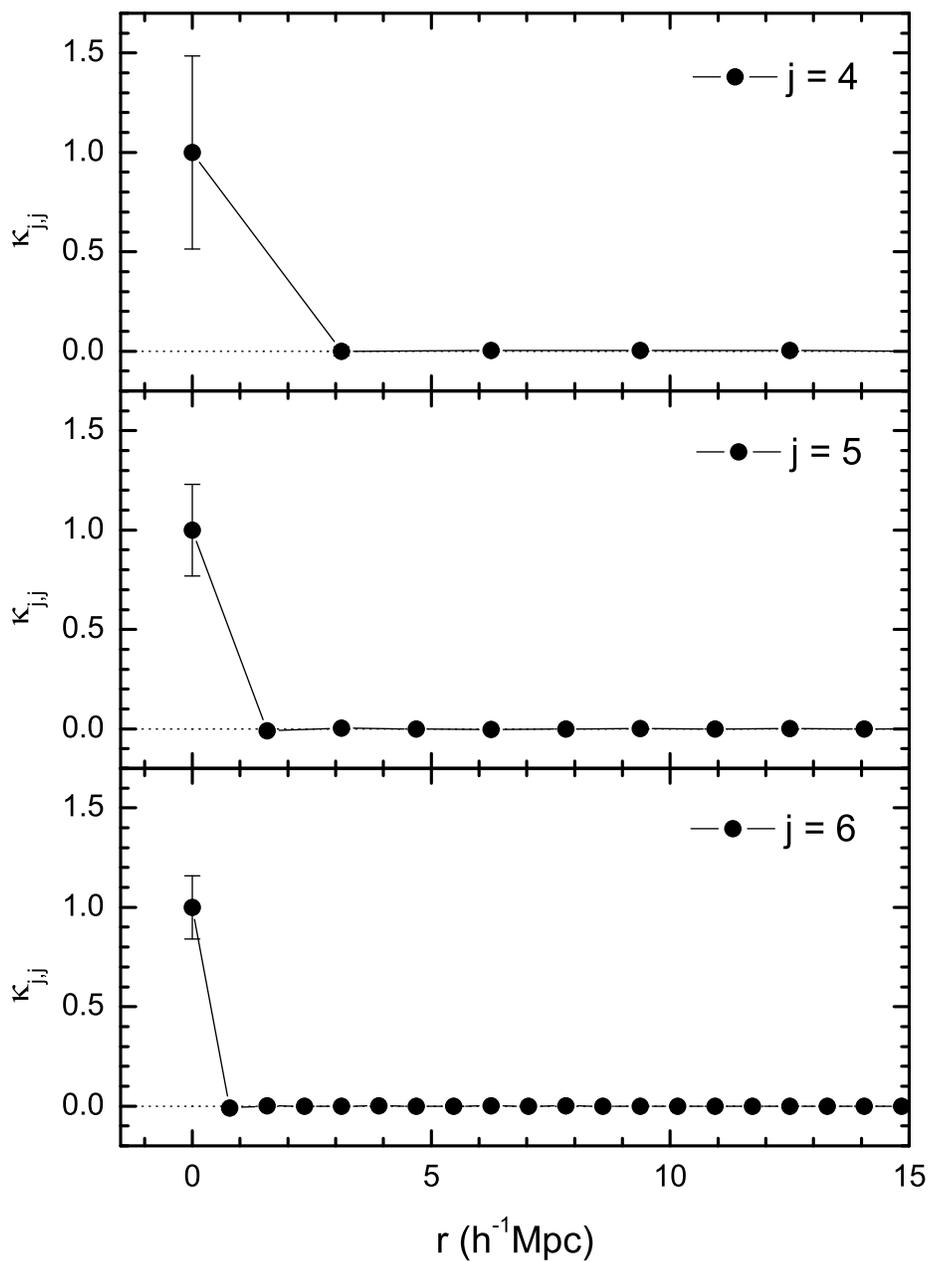}\caption{$\kappa_{\bf
j,j}$ vs. $r$ for the simulation data. ${\bf j}$ are taken to be
$(4,4,4)$ (top panel), (5,5,5) (middle), and (6,6,6) (bottom). The
physical distance is given by $r=|{\bf l-l'}|50/2^j$ h$^{-1}$ Mpc.
}
\end{figure}

\begin{figure}
\figurenum{4}\epsscale{0.85}\plotone{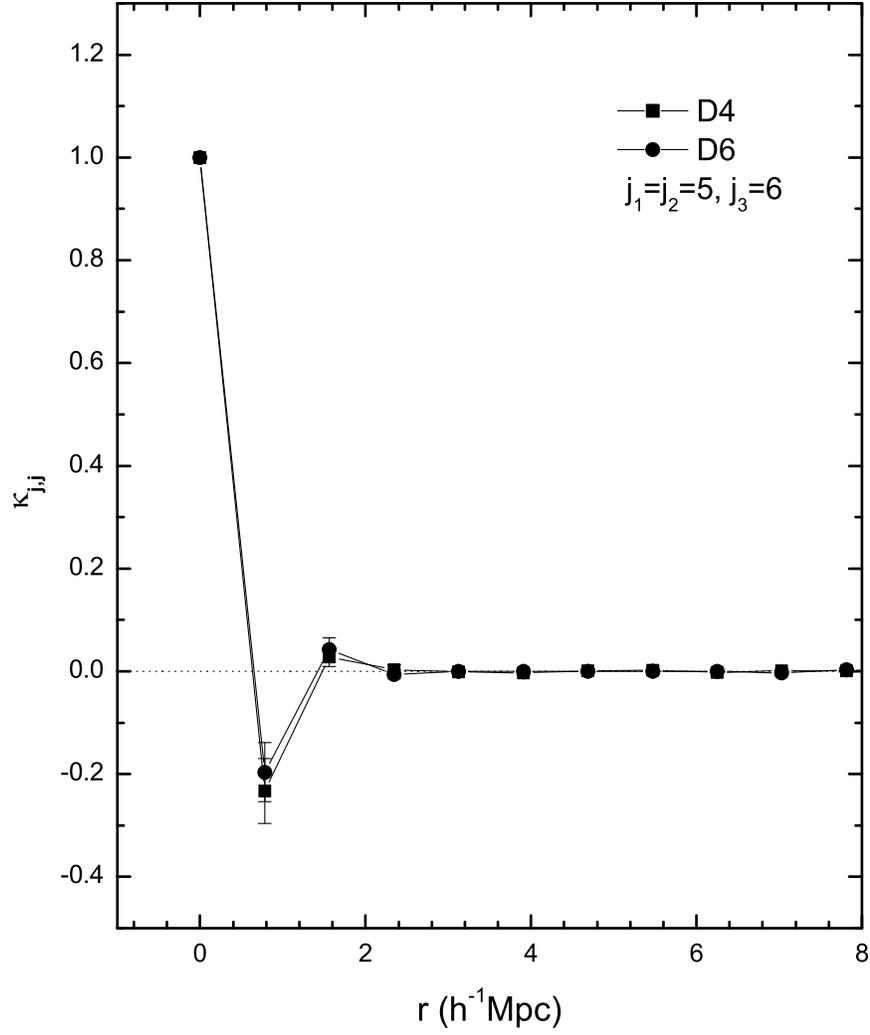}\caption{$\kappa_{\bf
j,j'}(\Delta l)$ vs. $r$ for the simulation data. The scales are
${\bf j}=(5,5,6)$ The physical distance $r=|\Delta {\bf l}| 50/
2^{6}$ h$^{-1}$ Mpc. Wavelets D4($M=2$) and D6($M=3$) are used. }
\end{figure}

\begin{figure}
\figurenum{5}\epsscale{0.85}\plotone{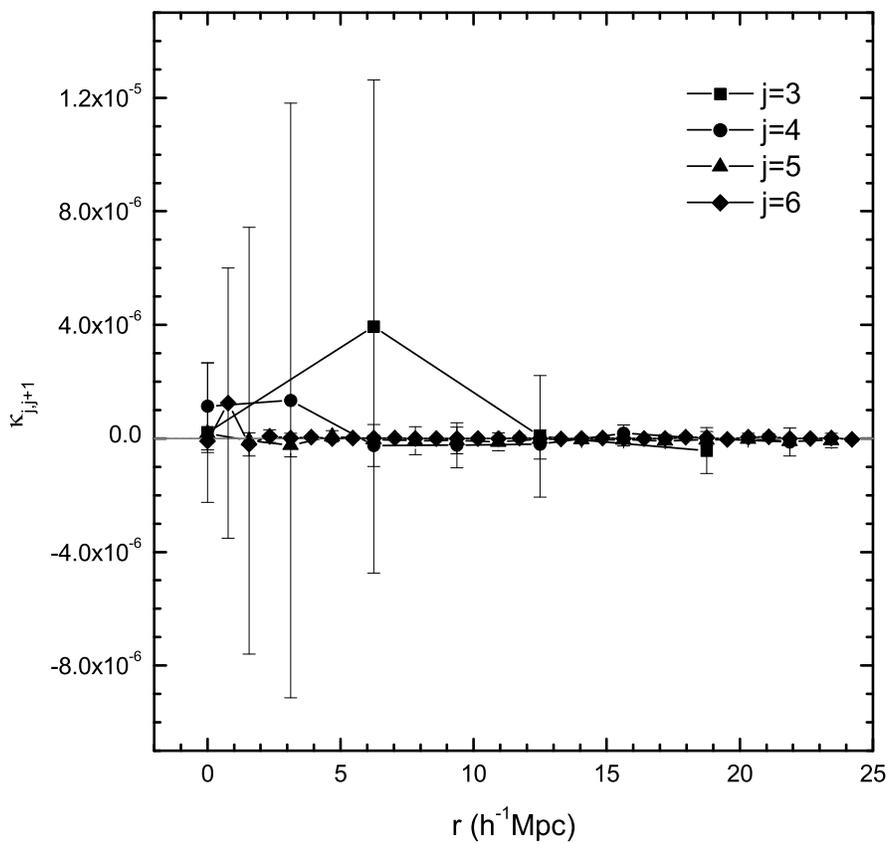}\caption{$\kappa_{\bf
j,j+1}(\Delta {\bf l})$ vs. $r$ for the simulation data. The scale
indexes are taken to be ${\bf j}=(j,j,j,)$, ${\bf
j'}=(j+1,j+1,j+1)$, and $j=3$, 4, 5 and 6, where $\Delta{\bf
l}={\bf l}- {\bf l'}/2$. The physical distance $r=|\Delta{\bf l}|
50/2^{j}$ h$^{-1}$ Mpc. }
\end{figure}

\begin{figure}
\figurenum{6}\epsscale{0.85}\plotone{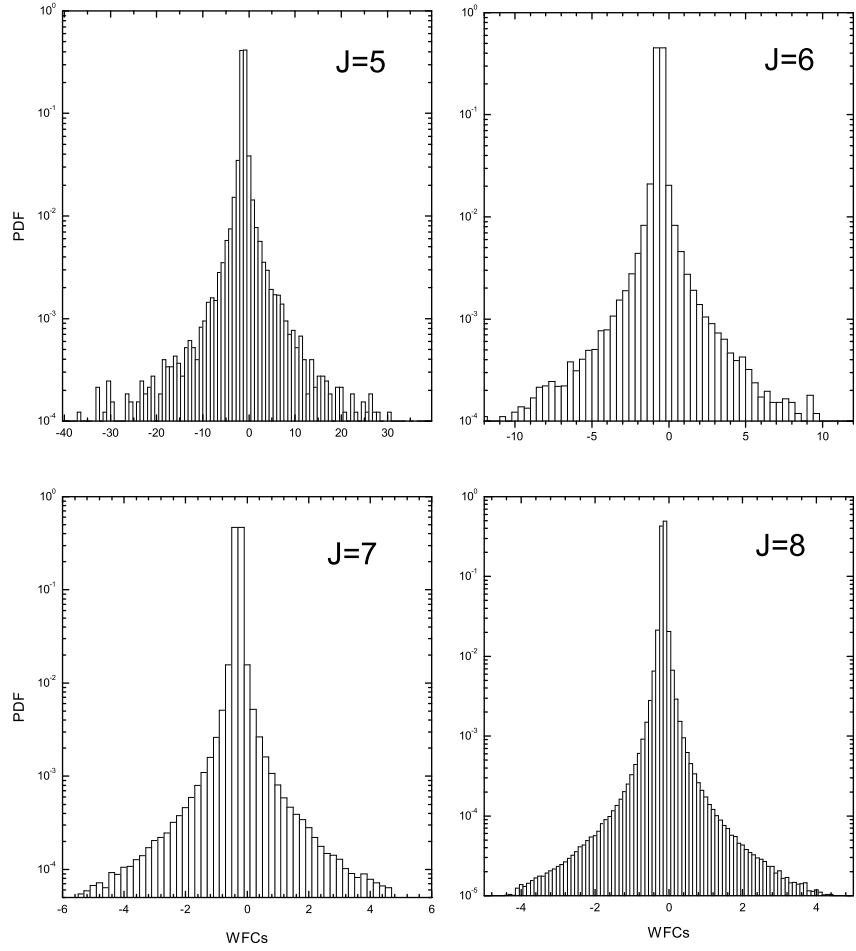}\caption{The
one-point distribution of $\tilde{\epsilon}_{\bf j, l}$ for ${\bf
j}=(j,j,j,)$ and $j =$ 5, 6, 7 and 8. }
\end{figure}

\begin{figure}
\figurenum{7}\epsscale{0.85}\plotone{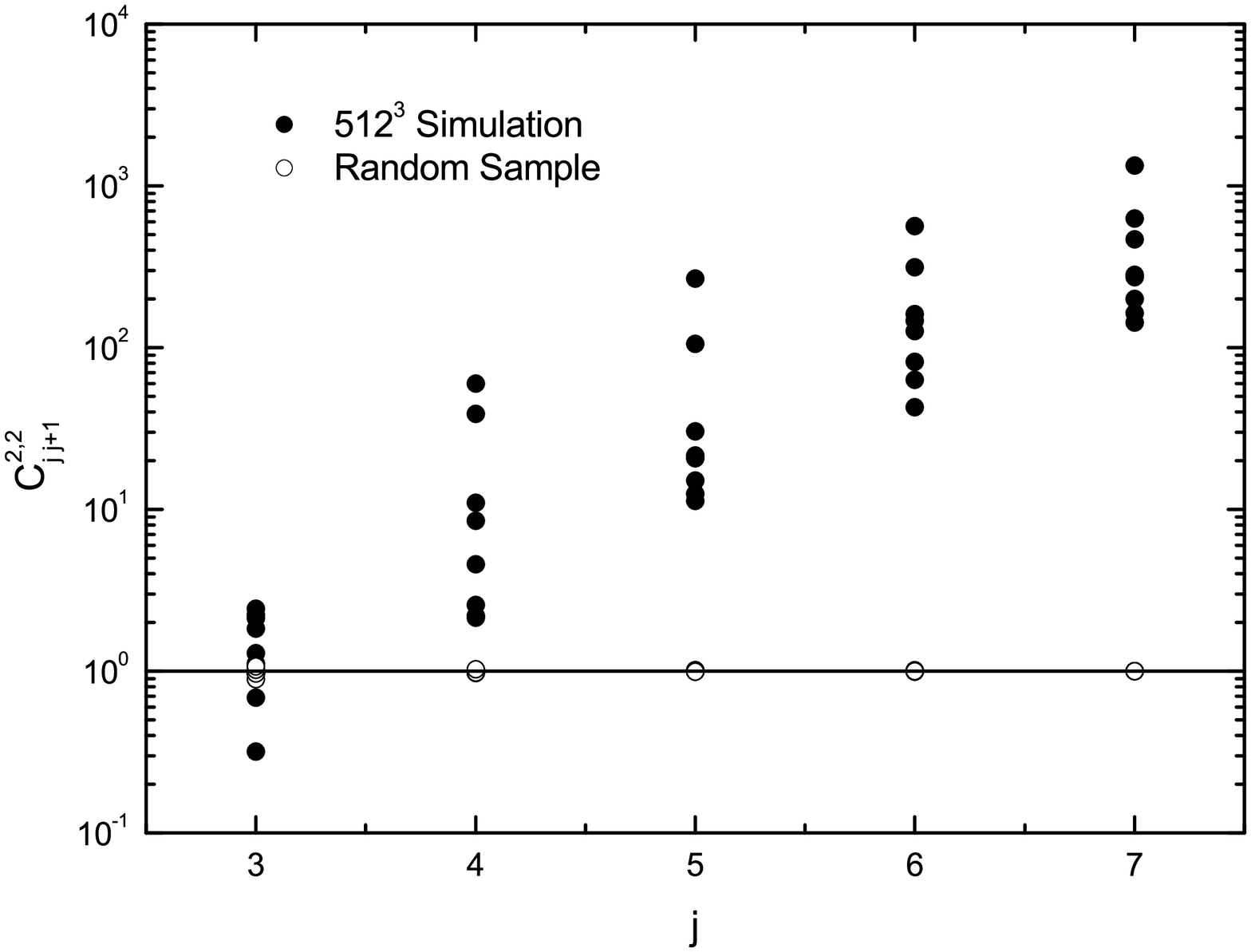} \caption{The
4$^{th}$ local scale-scale correlation $C^{2,2}_{\bf j,j+1}(\Delta
{\bf l}=0)$ vs. $j$ for the simulation data (solid circle) and
Gaussian random sample (circle). }
\end{figure}


\begin{references}

\reference{} Bardeen, J., Bond, J., Kaiser, N. \& Szalay, A. 1986,
\apj, 304, 15

\reference{} Bi, H.G. 1993, \apj, 405, 479

\reference{} Bi, H.G \& Davidsen, A. F. 1997, \apj, 479, 523.

\reference{} Cole S. and Kaiser, N. 1988, \mnras, 233, 637

\reference{} Cooray, A. and Sheth, R. 2002, Physics Reports, 372, 1latex

\reference{} Croft, R.A.C., Weinberg, D., Katz, N. \&  Hernquist,
  L.  1998, \apj, 495, 44

\reference{} Daubechies I. 1992, {\it Ten Lectures on Wavelets}
 (Philadelphia: SIAM)

\reference{} Fang, L.Z. \& Feng, L.L. 2000, \apj,  539, 5

\reference{} Fang, L.Z. and Thews, R. 1998, Wavelets in Physics,
  World Scientific, (Singapore)

\reference{} Feng, L.L., Deng, Z.G. \& Fang, L.Z. 2000, \apj, 530, 53

\reference{} Feng, L.L. \& Fang, L.Z. 2000, \apj,  535, 519

\reference{} Flandrin, P., 1992, IEEE Trans. Inf. Theory, 1992, 38, 910

\reference{} Greiner, M., Eggers, H.C. \& Lipa, P. 1998, Phys. Rev. Lett.
   80, 5333

\reference{} Greiner, M., Giesemann, J., Lipa, P., and Carruthers, P. 1996,
  Z. Phys. C69, 305

\reference{} Greiner, M., Lip, P., \& Carruthers, P., 1995,
 Phys. Rev. E51, 1948.

\reference{} Jing, Y.P., \& Suto, Y., 2002, \apj, 574, 538

\reference{} Jones, B.T., 1999, \mnras, 307, 376

\reference{} Meneveau, C. \& Sreenivasan, K.R. 1987, Phys. Rev. Lett.
   59, 1424

\reference{} Monaco, P. \& Efstathiou, G. 2000, \mnras,308,763

\reference{} Moore, B., Governato, F., Quinn, T., Stadel, J. \&
      Lake, G., 1999, \mnras, 261, 827

\reference{} Narayanan, V. \&  Weinberg, D. 1998, \apj, 508, 440

\reference{} Navarro, J., Frenk, C. \& White, S. 1996, \apj, 462, 563

\reference{} Neyman, J. \& Scott, E.L.  1952, \apj, 116, 144

\reference{} Pando, J. \& Fang, L.Z. 1998, A\&A, 340, 335

\reference{} Pando, J. Feng, L.L.  \& Fang, L.Z. 2001, \apj, 554, 841

\reference{} Pando, J., Lipa, P., Greiner, M. and Fang, L.Z. 1998,
     \apj, 496, 9

\reference{} Press, W.H. \& Schechter, P. 1974, \apj, 187, 425

\reference{} Peebles, P. 1980, The large scale structures of the universe,
    (Princeton press)

\reference{} Ramanathan, J. \& Zeitouni, O.  1991, IEEE Trans. Inf. Theory,
   37, 1156

\reference{} Rodrigues, D.D.C. \& Thomas, P.A. 1996, \mnras 282, 631

\reference{} Scherrer, R.J. \& Bertschinger, E. 1991, \apj, 381, 349

\reference{} Soneira, R.M. \& Peebles, P.J.E. 1977, \apj, 211, 1

\reference{} Spergel, D.N. et al 2003, astro-ph/0302209

\reference{} Tewfik, A.H. and Kim, M. 1992, IEEE Trans. Inf. Theory,
  38, 904

\reference{} Zel'dovich, Ya.B. 1970, A \& A, 5, 84

\end{references}
\end{document}